# RENORMALISATION AND HIERARCHIES

S.B.M. Bell[†] and B.M. Diaz[‡]

Interdisciplinary Quantum Group, Department of Computer Science

The University of Liverpool, Chadwick Building, Peach Street

Liverpool, L69 7ZF, United Kingdom

## ABSTRACT

We describe a new method of calculating the renormalised energy of a field obeying the Maxwell and Dirac equations. The method does not involve evaluating integrals but relies instead on summing a geometric series. We show that the new solution is equivalent to summing over the levels in the gravitational hierarchies explored previously by Bell and Diaz [2004a]. We augment our gravitational model by the introduction of a shell composed of a smooth mixture of negative and positive energy. This separates the charge from the mass of a body and leads to a more general hierarchy. Summing over the levels in this then produces the renormalisation series found above for the electromagnetic field, as well as the renormalisation series required by gravity.

[†] Mail@sarahbell.org.uk

[‡] B.M.Diaz@csc.liv.ac.uk

1. **PRELIMINARIES**

1.1 **Note on Nomenclature**

Matrices consisting of more than one row or column, quaternions and maps are given boldface type. $^T$ signifies transposition. $^\dagger$ signifies Hermitian conjugation. $^\ddagger$ signifies quaternion conjugation [Altmann 1986]. A lowercase Latin subscript stands for 1, 2 or 3 and indicates the space axes. A lowercase Greek subscript stands for 0, 1, 2, or 3 and indicates the space-time axes. $i$ is the square root of $-1$. $\mathbf{i}_0 = 1$. $\mathbf{i}_1 = \mathbf{i}$, $\mathbf{i}_2 = \mathbf{j}$, $\mathbf{i}_3 = \mathbf{k}$ stand for the quaternion matrices, where $\mathbf{i}_r^2 = -1$ and $\mathbf{i}_1\mathbf{i}_2 = \mathbf{i}_3$, $\mathbf{i}_2\mathbf{i}_1 = -\mathbf{i}_3$ with cyclic variations [Altmann 1986]. We use $-i\boldsymbol{\sigma}_r$ as the basis for the quaternions, where $\boldsymbol{\sigma}_r$ are the Pauli matrices [Dirac 1958],

$$\boldsymbol{\sigma}_1 = \begin{pmatrix} 0 & 1 \\ 1 & 0 \end{pmatrix}, \quad \boldsymbol{\sigma}_2 = \begin{pmatrix} 0 & -i \\ i & 0 \end{pmatrix}, \quad \boldsymbol{\sigma}_3 = \begin{pmatrix} 1 & 0 \\ 0 & -1 \end{pmatrix}$$

Quaternion conjugation and Hermitian conjugation are therefore related by $(\mathbf{Q} + i\mathbf{U})^\ddagger = \mathbf{Q}^\dagger + i\mathbf{U}^\dagger$ where $\mathbf{Q}$ and $\mathbf{U}$ are real quaternions. For any quaternion, $\mathbf{Q} = q_0 + q_1\mathbf{i}_1 + q_2\mathbf{i}_2 + q_3\mathbf{i}_3$, where the $q_\mu$ may be complex, we call $q_0$ the temporal part and $q_1\mathbf{i}_1 + q_2\mathbf{i}_2 + q_3\mathbf{i}_3$ the spatial part. We extend this to expressions involving quaternions in the obvious way. We define the underline symbol and the vertical bar as indicating *a reflector* or *a rotator* matrix, respectively [Bell et al. 2000],

$$\underline{\mathbf{U}} = \underline{\mathbf{U}}(\mathbf{Q},\mathbf{U}) = \begin{pmatrix} 0 & \mathbf{Q} \\ \mathbf{U} & 0 \end{pmatrix}, \quad \mathbf{U}| = \mathbf{U}|(\mathbf{Q},\mathbf{U}) = \begin{pmatrix} \mathbf{Q} & 0 \\ 0 & \mathbf{U} \end{pmatrix}$$

where **Q** and **U** are quaternions, not necessarily real. $^T$ signifies *reflector transposition*, $\underline{\mathbf{U}}^T(\mathbf{Q},\mathbf{U}) = \underline{\mathbf{U}}^T(\mathbf{Q}^T,\mathbf{U}^T)$. We use units in which $c = h_g/2\pi = 1$, where $c$ is the speed of light and $h_g$ is a generalisation of Planck's constant, where appropriate, and drop the explicit mention of a charge where its presence is normally understood. Repetition of a subscript does not imply summation which is always signalled by the explicit use of $\sum$.

## 1.2 The Dirac and Maxwell Equations

Any solution of the Dirac equation is also a solution of *the trim Dirac equation* [Bell and Diaz 2004b],

$$\left(\underline{\mathbf{D}}(\mathbf{D},\mathbf{D}^\ddagger) - \underline{\mathbf{A}}(\mathbf{A},\mathbf{A}^\ddagger)\right)\underline{\Phi}(\phi_1,\phi_2) = \underline{\Phi}\,\underline{\mathbf{M}}(\mathbf{M},-\mathbf{M}^\ddagger) \qquad (1.2.\text{A})$$

We define *the bijection* **H** between vectors in $\mathfrak{I}^4$, the field of complex numbers, and quaternions,

$$\mathbf{H}(v_0,v_1,v_2,v_3) \rightarrow (v_0 + \mathbf{i}_1 v_1 + \mathbf{i}_2 v_2 + \mathbf{i}_3 v_3) \qquad (1.2.\text{B})$$

We call the left-hand-side *the vector associated with the quaternion* on the right-hand side. We may then define,

$$\begin{aligned}
\mathbf{D} &= \mathbf{H}\{(\partial/\partial x_0,\ \partial/\partial x_1,\ \partial/\partial x_2,\ \partial/\partial x_3)\} \\
\mathbf{A} &= \mathbf{A}_{(t)} + \mathbf{A}_{(s)} \\
\mathbf{A}_{(t)} &= A_0(A_1\mathbf{i}_1 + A_2\mathbf{i}_2 + A_3\mathbf{i}_2)/|A_1\mathbf{i}_1 + A_2\mathbf{i}_2 + A_3\mathbf{i}_2| \\
\mathbf{A}_{(s)} &= -|A_1\mathbf{i}_1 + A_2\mathbf{i}_2 + A_3\mathbf{i}_2| \\
\mathbf{M} &= m
\end{aligned} \qquad (1.2.\text{C})$$

where $m$ is the mass of the particle and $(A_0, A_1, A_2, A_3)$ is the potential. Bell and Diaz [2004b] should be consulted for an explanation of why it is necessary map the temporal co-ordinate of the potential to space and the three-vector part to time. $\underline{\Phi}(\phi_1, \phi_2)$ is the wave function, with quaternion elements, and it is related to Dirac's wave function as follows [Bell et al. 2000]. We may define a map, **F**, of a quaternion, **Q**, onto a column matrix with two complex components,

$$\mathbf{F}(\mathbf{Q}) = \mathbf{Q}\begin{pmatrix} 1 \\ 0 \end{pmatrix} \quad (1.2.\text{D})$$

Using this we find,

$$\mathbf{F}(\phi_1) \to \phi_1, \quad \mathbf{F}(\phi_2) \to \phi_2 \quad (1.2.\text{E})$$

and then,

$$\begin{pmatrix} \phi_1 \\ \phi_2 \end{pmatrix} = \begin{pmatrix} i & -i \\ 1 & 1 \end{pmatrix} \begin{pmatrix} \psi_1 \\ \psi_2 \end{pmatrix} \quad (1.2.\text{F})$$

where the trailing vector is Dirac's wave function. We may return to the more usual expression of Dirac's equation by taking the spatial part of **A** to be $\mathbf{A}_{(s)}$ and the temporal part $\mathbf{A}_{(t)}$, and multiplying the temporal parts of $\underline{\mathbf{D}}$, $\underline{\mathbf{A}}$ and $\underline{\mathbf{M}}$ by $-i$. Finally, we set $\mathbf{A}_{(t)} = A_0$ and $\mathbf{A}_{(s)} = A_1\mathbf{i}_1 + A_2\mathbf{i}_2 + A_3\mathbf{i}_3$. For a full discussion see Bell and Diaz [2004b]. Our remarks below apply to both the new and old versions of the Dirac equation, as may be ascertained by applying our remarks on one to the other, and checking that the results still follow.

The Dirac equation, (1.2.A), has a conserved current, $\mathbf{H}^{-1}(\mathbf{J})$, [Bell and Diaz 2004b], given by,

$$\underline{\mathbf{J}}(\mathbf{J}, \mathbf{J}^{\ddagger}) = \underline{\Phi}^{\ddagger\text{T}} \underline{\Phi} \underline{\mathbf{K}}(\mathbf{k}, \mathbf{k}^{\ddagger}) \quad (1.2.\text{G})$$

where $\mathbf{k} = 1$ in the frame where $\underline{\phi}_1 = \underline{\phi}_2$. In turn this leads to *the trim radiation equation* in the Lorentz gauge,

$$\underline{\mathbf{DD}}\,\underline{\mathbf{A}} = \underline{\mathbf{J}} \qquad (1.2.\text{H})$$

We must impose the same structure on the current, $\mathbf{J}$, and $\mathbf{k}$ as we did on the potential in equations (1.2.C). Alternatively, we may revert to the more usual expressions, as we did for the potential, and multiply the temporal parts of $\underline{\mathbf{D}}$, $\underline{\mathbf{A}}$, $\underline{\mathbf{J}}$ and $\underline{\mathbf{K}}$ by $-i$. Our remarks below apply to both the new and old versions, and we will call both Maxwell's equations.

We may solve the Dirac equation in a constant potential to yield either a solution to the original Dirac equation [Bell and Diaz 2004b] or,

$$\underline{\Phi}(\underline{\phi}_1, \underline{\phi}_2) = \underline{\mathbf{Y}}(\mathbf{Y}_1, \mathbf{Y}_2) \exp\{\mathbf{i}_r(\nu x_0 + \mu_r x_r)\}, \qquad (1.2.\text{I})$$
$$\mathbf{Y}_2 = \mathbf{Y}_1\{-\mathbf{i}_r \nu + \mu_r + \mathbf{A}\}\mathbf{M}/m^2$$

where $\nu$ is the frequency and $\mu_r$ the wave number. This also determines the matching solution of the free Dirac equation, $\underline{\Phi}_0$.

$$\underline{\Phi}_0 = \underline{\mathbf{Y}}(\mathbf{Y}_1, \mathbf{Y}_2) \exp\{\mathbf{i}_r(\eta x_0 + \mu_r x_r)\}, \qquad (1.2.\text{J})$$
$$\mathbf{Y}_2 = \mathbf{Y}_1\{-\mathbf{i}_r \eta + \mu_r\}\mathbf{M}/m^2$$

$\nu$ and $\eta$ are related by [Bell and Diaz 2004b],

$$\nu = \eta + Z_p e A_0 \qquad (1.2.\text{K})$$

where we have indicated the magnitude of the charge on the particle, $Z_p e$, explicitly. If we apply this solution to the electron, $e$ is the magnitude of the charge on it, as it will prove convenient to take $A_0$ to be negative and the bound state has less energy than the free electron.

## 1.3 Bohr's Equations

The Maxwell and Dirac equations describe the behaviour of two types of radiation and associated particles [Bell and Diaz 2004b]. The first type, called *terrestrial*, travels at the speed of light, while the second, called *celestial*, travels infinitely fast between bodies at rest with respect to each other. The difference between the two lies in the way the equations transform between frames. In the terrestrial case, Lorentz transformations apply. In the celestial, rotations between the temporal and spatial co-ordinates using a real angle are appropriate.

The two-body interaction in which the Dirac equation describes the behaviour of a point body feeling an inverse distance potential from a point source, in the rest frame of the source, can be solved in the manner Dirac prescribed [Dirac 1958]. The scalar potential is the same in the terrestrial and celestial case [Bell and Diaz 2004b], nor is there is no overt mention of any change of frame here. We therefore expect the same solution in the celestial case, since the same trim Maxwell and Dirac equations can be used. This means that our new derivation of Bohr's equations from the Maxwell and Dirac equations [Bell et al. 2004a&b] must be re-interpreted in the celestial case, because Bohr's equations were derived in a space $M$ using the Lorentz transformations appropriate for the terrestrial case. $M$ arose when the Maxwell and Dirac equations were transformed from the space we called $L$, where Dirac's solution held [Bell et al. 2004a&b], [Bell 2004], into a new form in a new space, $M$. For the celestial case the factors in the Lorentz transformation become [Bell and Diaz 2004b],

$$\frac{1}{\sqrt{1-v^2}} \to \sqrt{1-v^2}, \quad \frac{v}{\sqrt{1-v^2}} \to v \qquad (1.3.A)$$

where *v* is the velocity of the Bohr particle and we use the Pythagorean theorem to find the magnitude of a four-vector in the celestial case. It is easily demonstrated by following the original derivation [Bell et al. 2004a] that when these substitutions are made Bohr's first equation remains unchanged,

$$\frac{Z_p Z_s e^2}{r} = \frac{mv^2}{\sqrt{1-v^2}}, \qquad (1.3.B)$$

where *r* is the Bohr radius, $Z_s e$ is the charge on the source and $n_\theta$ is an integer, with the proviso that charges with the opposite sign attract for the terrestrial version and charges with the same sign attract for the celestial version. We remark further on this below in section 2.2. Bohr's second equation depends entirely on the frame in which the quantisation is performed and on what. If we wish to reproduce the terrestrial version celestially,

$$\frac{mvr}{\sqrt{1-v^2}} = n_\theta \frac{h_g}{2\pi} \qquad (1.3.C)$$

we will need to assume that we are quantising a circular motion of the system as a whole round the particle in the rest frame of the particle. This follows because the energy of the system as a whole changes from the terrestrial version, $m\sqrt{1-v^2}$, to the celestial version, $m/\sqrt{1-v^2}$, from the relations (1.3.A). The spatial component of the latter is obtained by multiplying by the velocity, *v*, from the same relations. We then insist that the spatial part of the wave function of the system as a whole is single-valued, which produces Bohr's second equation in the same way as it did in the terrestrial case [Bell et al. 2004a&b]. We may re-interpret Bohr's first equation in the same way as a circular motion of the system as a whole in the rest frame of the particle. We may find the unrenormalised velocity, $v$, from equations (1.3.B) and (1.3.C),

$$\tilde{\alpha} = \frac{2\pi Z_s Z_p e^2}{n_\theta} = \frac{Z_s Z_p \alpha}{n_\theta} \qquad (1.3.D)$$

where $\alpha$ is the fine structure constant.

## 2. THE RENORMALISED SOLUTION

### 2.1 Derivation for a Constant Potential

We drop explicit mention of the charges in this section. We enhance the unrenormalised solution of Bohr's equations, (1.3.B), to provide a solution to all orders, initially by following Greiner and Reinhardt's [1996] derivation. Let the free Dirac equation in $L$ be given by,

$$\underline{\mathbf{D}}\,\underline{\boldsymbol{\Phi}}_0(t) - \underline{\boldsymbol{\Phi}}_0(t)\underline{\mathbf{M}} = 0 \qquad (2.1.A)$$

where we indicate the dependence of the wave function on the temporal co-ordinate explicitly, $x_0 = t$. We suppose that a radial inverse distance potential, is switched on suddenly for a short time, $\delta t_1$, at time $x_0 = t_1$. This causes the space $M$ to form and the Dirac equation in $M$ becomes [Bell et al. 2004a&b]

$$\underline{\mathbf{D}}\,\underline{\boldsymbol{\Phi}}_1(t_1) - \underline{\boldsymbol{\Phi}}_1(t_1)\underline{\mathbf{M}} = \underline{\mathbf{A}}\,\underline{\boldsymbol{\Phi}}_1 \qquad (2.1.B)$$

where $\underline{\mathbf{A}}$ is constant and we are only interested in the temporal part of $\underline{\mathbf{D}}$, which does not change. Since $\underline{\mathbf{A}}$ is assumed small, we may set,

$$\underline{\mathbf{\Phi}}_1(t_1) = \underline{\mathbf{\Phi}}_0(t_1) + \delta\underline{\mathbf{\Phi}}_1(t_1) \tag{2.1.C}$$

We obtain from equation (2.1.A) and (2.1.B) to first order,

$$\underline{\mathbf{D}}\delta\underline{\mathbf{\Phi}}_1(t_1) - \delta\underline{\mathbf{\Phi}}_1(t_1)\underline{\mathbf{M}} = \underline{\mathbf{A}}\,\underline{\mathbf{\Phi}}_0(t_1) \tag{2.1.D}$$

Integrating in the time interval $t_1$ to $t_1 + \delta t_1$ when the potential is acting,

$$\underline{\mathbf{I}}(1,1`)\delta\underline{\mathbf{\Phi}}_1(t_1 + \delta t_1) = \tag{2.1.E}$$

$$\int_{t_1}^{t_1+\delta t_1}\{-\underline{\mathbf{D}}_{(s)}\delta\underline{\mathbf{\Phi}}_1(t_1)\mathrm{d}t_1 + \delta\underline{\mathbf{\Phi}}_1(t_1)\underline{\mathbf{M}}\mathrm{d}t_1\} +$$

$$\int_{t_1}^{t_1+\delta t_1}\underline{\mathbf{A}}\,\underline{\mathbf{\Phi}}_0(t_0)\mathrm{d}t_1$$

where $\underline{\mathbf{D}}_{(s)}$ indicates the spatial part of $\underline{\mathbf{D}}$. The terms inside the curly brackets are of second order and we obtain an estimate of the scattered waves for a purely temporal potential, $\mathbf{A}_{(t)}$,

$$\frac{\delta\underline{\mathbf{\Phi}}_1(t_1 + \delta t_1)}{\delta t_1} = \mathbf{A}\underline{\mathbf{\Phi}}_0(t_1) \tag{2.1.F}$$

We use the approximate expression for a differential of the wave function, $\underline{\mathbf{\Phi}}_1$, to obtain an estimate for the full change of gradient of both the original and scattered wave between $t_1$ and $t_1 + \delta t_1$. We will call the real function that describes the behaviour of the sum of both waves $\mathbf{\Phi}^{(1)}$, then,

$$\frac{\delta\underline{\mathbf{\Phi}}^{(1)}(t_1)}{\delta t_1} = \frac{\underline{\mathbf{\Phi}}_0(t_1 + \delta t_1) - \underline{\mathbf{\Phi}}_0(t_1)}{\delta t_1} + \frac{\delta\underline{\mathbf{\Phi}}_1(t_1 + \delta t_1)}{\delta t_1} \tag{2.1.G}$$

Approximating the discrete differential as a continuous function,

$$\frac{\delta \underline{\Phi}^{(1)}(t_1)}{\delta t_1} \approx \frac{\partial \underline{\Phi}_0(t_1)}{\partial t_1} + \frac{\delta \underline{\Phi}_1(t_1 + \delta t_1)}{\delta t_1} \qquad (2.1.\text{H})$$

Using equations (1.2.J) and (2.1.F) this becomes,

$$\frac{\delta \underline{\Phi}^{(1)}(t_1)}{\delta t_1} \approx (\mathbf{i}_r \eta + \mathbf{A}) \underline{Y} \exp\{\mathbf{i}_r (\eta t_1 + \mu_r x_r)\} \qquad (2.1.\text{I})$$

where we will assume the potential, $\mathbf{A} = \mathbf{i}_r A_0$, lies along $\mathbf{i}_r$ [Bell and Diaz 2004b] and that $x_r$ always lies in the relevant space. Since $\mathbf{i}_r$ has been substituted for the $i$ in our familiar solution to the original Dirac equation, the energy operator becomes $-\mathbf{i}_r \partial/\partial x_0$ and we may see that $\eta + A_0$ has been substituted for the original eigenvalue, $\eta$ by replacing $\partial \underline{\Phi}^{(1)}/\partial t_1$ by $\delta \underline{\Phi}^{(1)}/\delta t_1$. We are thus required to assign an approximate value for the total wave function of,

$$\underline{\Phi}^{(1)}(t_1) = \underline{Y} \exp\{\mathbf{i}_r [(\eta + A_0)t_1 + \mu_r x_r]\} \qquad (2.1.\text{J})$$

This is the same as the unrenormalised solution obtained by solving the Dirac equation (1.2.A) from equations (1.2.I) and (1.2.K).

The usual next move in the calculation of the series by this route is to switch the potential off and consider the development of $\delta \underline{\Phi}_1(t_1 + \delta t_1)$ according to the free propagator, which leads to the Feynman series. However, we gain nothing from propagating the scattered wave here, since we do not have to allow the potential to vary with location and time and our method is not sensitive to differences in the phase of the propagated wave function. Nor do we want to switch the potential off for this bound state. So we will pick the instant $t_1 + \delta t_1$ as the point the potential acts again. The particle is going in a straight line in *M*, taking the radius and infinitesimal arc as Cartesian co-ordinates, as implied by the Dirac equation in *M* [Bell et al. 2004a] and so we may iterate. A new space is formed, which we shall call $M^2$.

Setting $t_2 = t_1 + \delta t_1$ and supposing the potential to act for a time $\delta t_2$, we may solve the Dirac equation again, to produce,

$$\frac{\delta \underline{\Phi}_2(t_2 + \delta t_2)}{\delta t_2} = \mathbf{A} \underline{\Phi}_1(t_2) \tag{2.1.K}$$

in an exact analogy with equation (2.1.F). We may then admit a further term into equation (2.1.G) for an improved estimate of the gradient of the total wave at $t_2$,

$$\frac{\delta \underline{\Phi}^{(2)}(t_2)}{\delta t_2} = \frac{\underline{\Phi}_0(t_2 + \delta t_2) - \underline{\Phi}(t_2)}{\delta t_2} + \frac{\delta \underline{\Phi}_1(t_2 + \delta t_2)}{\delta t_2} + \frac{\delta \underline{\Phi}_2(t_2 + \delta t_2)}{\delta t_2} \tag{2.1.L}$$

approximating,

$$\frac{\delta \underline{\Phi}^{(2)}(t_2)}{\delta t_2} \approx \frac{\partial \underline{\Phi}_0(t_1)}{\partial t_1} + \frac{\delta \underline{\Phi}_1(t_1 + \delta t_1)}{\delta t_1} + \frac{\delta \underline{\Phi}_2(t_2 + \delta t_2)}{\delta t_2} \tag{2.1.M}$$

and calculating the last term from equation (2.1.K),

$$\frac{\delta \underline{\Phi}_2(t_2 + \delta t_2)}{\delta t_2} \approx \mathbf{A} \int_{t_2}^{t_1} \left( \frac{\delta \underline{\Phi}_1(t_1 + \delta t_1)}{\delta t_1} \right) \mathrm{d} t_1 \tag{2.1.N}$$

Then from equation (2.1.F),

$$\frac{\delta \underline{\Phi}_2(t_2 + \delta t_2)}{\delta t_2} \approx \mathbf{A} \int_{t_2}^{t_1} (\mathbf{A} \underline{\Phi}_0(t_1)) \mathrm{d} t_1 \tag{2.1.O}$$

and finally we obtain from equation (1.2.J),

$$\frac{\delta \underline{\Phi}_2(t_2 + \delta t_2)}{\delta t_2} \approx -\mathbf{i}_r \frac{\mathbf{A}^2}{\eta} \underline{\Phi}_0(t_1) \tag{2.1.P}$$

We obtain an estimate of the eigenvalue and total wave function from equations (2.1.I), (2.1.J) and (2.1.M),

$$\frac{\delta \underline{\Phi}^{(2)}(t_1)}{\delta t_1} \approx \mathbf{i}_r \left( \eta + A_0 + \frac{A_0^2}{\eta} \right) \underline{Y} \exp\{\mathbf{i}_r (\eta t_1 + \mu_r x_r)\}, \tag{2.1.Q}$$

$$\underline{\Phi}^{(2)}(t_1) = \underline{Y} \exp\left\{ \mathbf{i}_r \left[ \left( \eta + A_0 + \frac{A_0^2}{\eta} \right) t_1 + \mu_r x_r \right] \right\}$$

It is clear we may iterate by having the potential act again in $M^2$ to create $M^3$ and that we may repeat this for $M^4, M^5 \ldots M^n$ so that,

$$\underline{\Phi}^{(n)}(t_1) = \tag{2.1.R}$$

$$\underline{Y} \exp\left\{ \mathbf{i}_r \left[ \eta \left( 1 + \frac{A_0}{\eta} + \left(\frac{A_0}{\eta}\right)^2 + \ldots + \left(\frac{A_0}{\eta}\right)^n \right) t_1 + \mu_r x_r \right] \right\}$$

The value of $\eta$ depends upon our chosen frame and circumstances. Supposing that $A_0 < \eta$ and summing the series to infinity,

$$\underline{\Phi}^{(\infty)} = \tag{2.1.S}$$

$$\underline{Y} \exp\left( \mathbf{i}_r \left[ \left\{ \eta + \frac{A_0}{1 - A_0/\eta} \right\} x_0 + \mu_r x_r \right] \right)$$

We may then produce the renormalised solution to the Dirac and Maxwell equations by replacing the potential in the unrenormalised solution by,

$$A_0 \to \frac{A_0}{1 - A_0/\eta} \tag{2.1.T}$$

from equation (1.2.K). We call the potential after this amendment *the adjusted potential*.

## 2.2 An Arbitrary Field

We return to our demonstration that the particle version of QED theory may be inferred from the information that the two-body interaction obeys Bohr's equations [Bell et al. 2004a&b]. We did this by showing that we could calculate a wave function for a particle feeling the effects of an arbitrary distribution of charge independently of the boundary conditions. We split the charge into infinitesimal discs or spheres and calculated the potential on the boundary or surface. Then we found the wave function of a particle feeling that potential and obeying the Bohr equations. This wave function satisfies the Dirac equation and we arranged for the surfaces of the discs or spheres to be linked so that they covered the space *L*, providing, on addition of the boundary conditions, a solution of the Dirac equation over *L*. We can renormalise by adjusting the potential arising from each disc or sphere using equation (2.1.T). This means that we must know $\eta$ and hence $v$. We discover these from the renormalised solution to the Bohr equations.

Bohr's first equation is derived from an inverse distance potential [Bell et al. 2004a&b] and to reflect the renormalised solution this should become the adjusted potential. We derived Bohr's first equation using equation (1.2.K) and de Broglie's relations to give us the kinetic energy, $\eta$, of the free electron and energy of the system as a whole, $v$. In the terrestrial case,

$$\eta = \frac{m}{\sqrt{1-v^2}}, \quad \nu = m\sqrt{1-v^2}, \tag{2.2.A}$$

In the celestial case the expressions for $\eta$ and $\nu$ are interchanged, but the resultant equation is the same because we change the sign of one of the charges. Upon substitution into equation (1.2.K) and the application of equation (2.1.T),

$$\mathbf{A}(v^2 - 1) = \frac{mv^2}{\sqrt{1-v^2}} \tag{2.2.B}$$

where $\mathbf{A}$ is the original potential. Since our potential in the previous section is not distinguished from the potential energy, given by $-Z_p Z_s e^2/r$ assuming that $Z_p$, $Z_s$ and $e$ are all positive,

$$\frac{Z_p Z_s e^2}{r} = \frac{mv^2}{\sqrt{1-v^2}(1-v^2)} \tag{2.2.C}$$

$Z_s e$ is assumed to be known since it is the total charge within the disc or sphere. Solving for the renormalised velocity, $v$, in terms of the unrenormalised velocity, $\tilde{\alpha}$, by using equation (1.3.C),

$$\tilde{\alpha}^2 v^2 + v - \tilde{\alpha} = 0 \equiv v = \frac{-1 \pm \sqrt{1+4\tilde{\alpha}^2}}{2\tilde{\alpha}} \tag{2.2.D}$$

The two solutions for small and large $\tilde{\alpha}$, respectively, are

$$v = \tilde{\alpha} + \mathrm{O}(\tilde{\alpha}^3), \quad v = -\frac{1}{\tilde{\alpha}} - \tilde{\alpha} + \mathrm{O}(\tilde{\alpha}^3), \quad v = \pm 1 - \frac{1}{2\tilde{\alpha}} + \mathrm{O}(\tilde{\alpha}^{-2}) \tag{2.2.E}$$

This is a surprising result. In addition to the expected answer, $v \approx \tilde{\alpha}$ where $\tilde{\alpha}$ is small, we get $v \approx -1/\tilde{\alpha}$, and, since $\tilde{\alpha}$ is still small, $|v|$ is large. For large $\tilde{\alpha}$ we find a third possibility, $v \approx 1$. We will comment on this after we have finished describing the renormalised solution in an arbitrary field.

We may now calculate the adjusted potential using equation (2.1.T) or equation (2.2.B). If this is used for the Dirac equation the result will be accurate to all orders, given our discussion in section 2.1. The correction could be summed up by deriving a new version of Maxwell's equations which delivered the adjusted potential but it would also be necessary to specify which particle is to be influenced, because its mass, charge and level of excitation, $n_\theta$, would necessarily appear. This new version of Maxwell's equations would therefore be a radical departure that we do not undertake.

Finally, it may be that some of the interactions we classify under different names, for example, quantum flavourdynamics or quantum chromodynamics, might be reachable using the alternative values for the fine structure constant given in equations (2.2.E) and the Dirac equation for new particles. We say this more particularly because of the triune nature of the electric charge in the trim Dirac equation [Bell and Diaz 2004b].

## 3.  THE CONNECTION WITH HIERARCHIES

### 3.1  The Unishell and Copyshell

We turn to the relationship between the series representing the energy of a system which we found in equation (2.1.R),

$$R_s = \eta\left(1 + \frac{A_0}{\eta} + \left(\frac{A_0}{\eta}\right)^2 + \ldots + \left(\frac{A_0}{\eta}\right)^n + \ldots + \right) = \frac{\eta^2}{\eta - A_0} \tag{3.1.A}$$

and the hierarchies of thin shells found for gravity in the terrestrial case [Bell and Diaz 2004a], [Bell 2004]. These hierarchies also apply in the celestial case, for two reasons. First, Bohr's equations, which are the same in both cases, specify the interaction at each level. Next, we define *the Dirac energy*. This is the unrenormalised energy found by the usual energy operator operating on a wave function obeying Dirac's equation with a potential obeying Maxwell's equations. Then, second, the rules for going from one level of a hierarchy to the next are the same, provided we suppose that, for both in the same way, the Dirac energy at one level is the origin of the charge on the source for the interaction at one level higher.

We obtained such a hierarchy for the unishell, a constant series that appears whenever we consider a body feeling the effect of gravity [Bell and Diaz 2004a].

**Table 1.**

**Part of the unishell hierarchy**

| **Level,** $r$ | $-1$ | $0$ | $1$ | $2$ | $3$ |
|---|---|---|---|---|---|
| **Source** $m_u^{(r)}$ | $\dfrac{m_u}{4}$ | $\dfrac{m_u}{2}$ | $m_u$ | $2m_u$ | $4m_u$ |
| **Velocity** $v^{(r)} = v_r$ | $v_{-1} = \dfrac{1}{2\sqrt{2}}$ | $v_0 = \dfrac{1}{2}$ | $v_1 = \dfrac{1}{\sqrt{2}}$ | $v_2 = 1$ | $v_3 = \sqrt{2}$ |
| **Energy** $E_u^{(r)}$ | $m\sqrt{7/8}$ | $m\sqrt{3/4}$ | $m/\sqrt{2}$ | $0$ | $mi$ |

where $m_u^{(r)}$ is the mass of the source, in this case the unishell, $v_r$ is the velocity of a particle in orbit round the unishell, $E_u^{(r)}$ is the Dirac energy, $m$ is the gravitational charge and mass of the particle, superscript $(r)$ always indicates level *r*. We have extended the hierarchy past the velocity of light, although Bohr's equations break down here, by simple iteration of the rules for going from one level to the next. We shall examine this more closely in sections 3.3 and 3.4. Table 1 could be written,

**Table 2.**

**Alternative expressions for the unishell hierarchy**

| Level | $r-2$ | $r-1$ | $r$ | $r+1$ | $r+2$ |
|---|---|---|---|---|---|
| **Source** $m_u^{(r)}$ | $m_u v_r^4$ | $m_u v_r^2$ | $m_u$ | $m_u v_r^{-2}$ | $m_u v_r^{-4}$ |
|  | $m_u v_{r-1}^2$ | $m_u v_r^2$ | $m_u v_{r+1}^2$ | $m_u v_{r+2}^2$ | $m_u v_{r+3}^2$ |
| **Velocity** $v^{(r)} = v_r$ | $v_{r-2} = \dfrac{1}{2\sqrt{2}}$ | $v_{r-1} = \dfrac{1}{2}$ | $v_r = \dfrac{1}{\sqrt{2}}$ | $v_{r+1} = 1$ | $v_{r+2} = \sqrt{2}$ |
|  | $v_r^3$ | $v_r^2$ | $v_r$ | $v_r^0$ | $v_r^{-1}$ |
| **Energy** $E_u^{(r)}$ | $m\sqrt{1-v_{r-2}^2}$ | $m\sqrt{1-v_{r-1}^2}$ | $m\sqrt{1-v_r^2}$ | $m\sqrt{1-v_{r+1}^2} = 0$ | $m\sqrt{1-v_{r+2}^2}$ |

where it becomes apparent that the hierarchy is independent of scale, for example, if we permit the velocity of light, *c*, to alter with the level, $c_r$. Moving from level *r* to level $r+1$ then requires, for $v_r/c_r$,

$$r \to r+1 \Rightarrow c_r \to \frac{c_r}{\sqrt{2}} \tag{3.1.B}$$

For Bohr's equations (1.3.B) and (1.3.C),

$$\frac{A_0}{\eta} = -v^2 \tag{3.1.C}$$

We could therefore write for the renormalised Dirac energy at the first level,

$$R_u^{(1)} = \frac{E_u^{(1)}}{(1-v_1^2)}(1 - v_1^2 + v_1^4 - v_1^6 + \ldots)$$
$$= \frac{E_u^{(1)}}{m_u(1-v_1^2)}\sum_{r=1}^{-\infty}(-1)^{r-1}m_u^{(r)} = \frac{E_u^{(1)}}{(1-v_1^2)^2}$$

(3.1.D)

from table 2 and equations (2.2.A) and (3.1.A). Renormalisation is now defined in terms of a sum over the hierarchy, which we will refine in section 3.4.

We will need to enhance the description of gravitational hierarchies we gave previously [Bell and Diaz 2004a] to allow the factor $\sqrt{2}$ which appeared ubiquitously there, for example in table 1, to take on other values, as happens in the renormalisation series in equation (3.1.A) applied to electromagnetism. We derived a more general hierarchy than the unishell, which we called the copyshell, and it is this description that we will generalise still further. The copyshell represents the combined effect of the unishell and some other shell, the copy shell. The latter could represent the earth, the solar system or some other gravitational body. The copyshell hierarchy is

**Table 3.**

**Part of the copyshell hierarchy**

| **Level** | -1 | 0 | 1 | 2 | 3 |
|---|---|---|---|---|---|
| **Source** $m_s^{(r)}$ | $m_s/4$ | $m_s/2$ | $m_s$ | $2m_s$ | $4m_s$ |
| **Particle** $m^{(r)} = m_r$ | $\dfrac{m_s/2}{\sqrt{1-v_g^2/4}}$ | $\dfrac{m_s}{\sqrt{1-v_g^2/2}}$ | $\dfrac{2m_s}{\sqrt{1-v_g^2}}$ | $\dfrac{4m_s}{\sqrt{1-2v_g^2}}$ | $\dfrac{8m_s}{\sqrt{1-4v_g^2}}$ |
| **Velocity** $v^{(r)}$ | $v_g/2$ | $v_g/\sqrt{2}$ | $v_g$ | $v_g\sqrt{2}$ | $2v_g$ |
| **Energy** $E_s^{(r)}$ | $m_{-1}\sqrt{1-v_g^2/4}$ | $m_0\sqrt{1-v_g^2/2}$ | $m_1\sqrt{1-v_g^2}$ | $m_2\sqrt{1-2v_g^2}$ | $m_3\sqrt{1-4v_g^2}$ |

where $m_s^{(r)}$ is the mass of the source, $m_r$ is the mass of the particle in orbit round it and $E_s^{(r)}$ is the Dirac energy.

### 3.2 Matter and Charge

We derived the unishell and copyshell hierarchies by showing that the curvature of space-time required by general relativity could also be represented by motion in a flat space-time [Bell 2004], [Bell and Diaz 2004a]. A closed circular path in the curved space-time was compared with circular motion in the flat space-time. These paths could be made arbitrarily small, at which point the similarity between the former and the path used for the derivation of the Riemann tensor is apparent [Martin 1995]. They differ only in the assumption of how many limits there are. There is only one if space is deemed continuous, usually reached by using the calculus, while there may be more, if a space has an hierarchical character. The difference can be illustrated with the difference between the a system at the critical point and the system away from criticality [Bell 1992]. In the first case, if we weight each tile or pixel equally, the size and shape of the clique do not matter, while in the second they affect the parameters which may be derived from a Markov model. If however, the cliques are given a discrete fractal pattern, rather than an identical weight, then there is no longer a single limiting state of criticality but many [Bell 1992]. The difference between the system on or off criticality becomes a difference of criticalities. In our current case, the continuous space of general relativity becomes a set of discrete steps away from a source at each of the Bohr radii for each level of the hierarchies [Bell and Diaz 2004a]. These radii were taken to be the same for both the flat and curved space and the interval in the former was set equal to the temporal co-ordinate in the latter. The angle turned in the former, $\hat{\theta}$, was related to the angle turned in the latter, $\theta$, by $\hat{\theta} = \theta\sqrt{2}$. We need to generalise this to,

$$\hat{\theta} = \theta\sqrt{2} \rightarrow \hat{\theta} = \theta/\chi \qquad (3.2.A)$$

where $\chi$ is an adjustable parameter, allowing us to join the motion at one level of the hierarchy smoothly onto the motion at the level below or above. We may introduce $\chi$ by inserting a permittivity into the metric.

For electromagnetism, if our units are such that the permeability is one, which we shall assume, we know that [Bleaney and Bleaney 1965],

$$c^2 = \frac{1}{\varepsilon} \qquad (3.2.B)$$

where $\varepsilon$ is the permittivity and $c$ is the velocity of an electromagnetic wave. The same function is performed by the gravitational constant, $G$, for gravity.

It would appear that we have no license to introduce a different permittivity into either electromagnetism or gravity, because *c* is the value found experimentally for the velocity of both light and gravitational waves. However, the change is temporary because we shall absorb this factor into the mass of the orbiting particle, once we have separated this from its charge, as we shall below. In practice, adding $\varepsilon \neq 1$ to our previous account [Bell and Diaz 2004a] can be achieved by replacing the mass of the source, $\tilde{M}_s$ by $\tilde{M}_s/\varepsilon$. To achieve the factor we wanted in equation (3.2.A), the permittivity must become

$$\varepsilon = \frac{1}{2\chi^2} \qquad (3.2.C)$$

since we found that $\theta \propto \sqrt{\tilde{M}_s}$ [Bell and Diaz 2004a].

We also need to introduce a more general parameter into our previous description of the self-gravitating thin spherical shell [Bell and Diaz 2004a], which we used to generate and calculate parameters for the copyshell. Here we supposed for the strength of the attraction,

$$F = \frac{2\tilde{M}_s^2}{\varepsilon r^2} \qquad (3.2.D)$$

where $2\tilde{M}_s$ was the mass of the shell, *r* its radius and we have added $\varepsilon$. The factor of a half was introduced because each infinitesimal portion attracts all the other portions except itself. We would like to generalise this by using a free parameter, $\beta$,

$$F = \frac{4\tilde{M}_s^2 \beta^2}{\varepsilon r^2} \qquad (3.2.E)$$

We may do this by supposing that the shell is composed of an even mixture of positive and negative mass in the ratio $1:k$. The positive mass will contribute a term proportional to $1/2$ as before. Since two negative masses leave the sign of *F* unchanged in equation (3.2.D), these will also attract, contributing $k^2/2$, while a positive and negative mass will repel each other, contributing $k$. We calculate the average attraction per unit squared of the material,

$$\beta^2 = \frac{1/2 + k^2/2 - k}{(1+k)^2} \tag{3.2.F}$$

This resolves to give the factor a half we found previously [Bell and Diaz 2004a] when $\tilde{k} = 0$ or in the limit as $\tilde{k} \to \infty$ and the shell is composed of all positive or negative mass. Simplifying,

$$\beta = \frac{1-k}{\sqrt{2}(1+k)} \tag{3.2.G}$$

where we have picked the positive square root. The equivalent of the particle must have a mass of,

$$m' = \frac{2\tilde{M}_s(1-k)}{1+k} \tag{3.2.H}$$

where we have subtracted the negative from the positive mass of the shell. From equation (3.2.G),

$$m' = 2\sqrt{2}\tilde{M}_s\beta \tag{3.2.I}$$

Bohr's first equation for the shell can now be derived following Bell and Diaz [2004a],

$$\frac{4\tilde{M}_s^2 \beta^2}{\varepsilon r} = \frac{m' v_s^2}{\sqrt{1 - v_s^2}}, \tag{3.2.J}$$

from equations (3.2.E) and where $v_s$ is the particle's velocity. This means that if we take the shell surface as the smeared-out wave function of the particle as we did before [Bell et al. 2004a&b], [Bell and Diaz 2004a], with $m'$ as the charge and mass, the equivalent of the source must have charge of $\sqrt{2}\tilde{M}_s \beta$, from equation (3.2.I). Following the method used by Bell and Diaz [2004a], Bohr's equations and associated definitions are now,

$$\frac{\sqrt{2}\tilde{M}_s \beta}{\varepsilon r} = \frac{v_s^2}{\sqrt{1 - v_s^2}}, \quad \frac{m'}{\sqrt{1 - v_s^2}} v_s r = n_\theta h'_g, \tag{3.2.K}$$

$$\frac{h'_g}{2\pi} = \frac{\sqrt{2}\tilde{M}_s \beta m'}{j\varepsilon}, \quad m' = 2\sqrt{2}\tilde{M}_s \beta, \quad v_s = \frac{j}{n_\theta}$$

where $j$ is an integer quantum number and $h'_g$ has the same role as Planck's constant. The second, third and fifth equations apply only if the motion is quantised. We absorb the permittivity into the mass, which is, as we have seen, a redefinition of the speed of the associated terrestrial radiation, and quote Bohr's equations as they would apply if we were to assume an electromagnetic interaction in the vacuum,

$$\frac{\sqrt{2}\tilde{M}_s \beta}{r} = \frac{\varepsilon v_s^2}{\sqrt{1 - v_s^2}}, \quad \frac{m}{\sqrt{1 - v_s^2}} v_s r = n_\theta h_g, \tag{3.2.L}$$

$$\frac{h_g}{2\pi} = \frac{\sqrt{2}\tilde{M}_s \beta m}{j\varepsilon}, \quad m = 2\sqrt{2}\tilde{M}_s \beta\varepsilon, \quad v_s = \frac{j}{n_\theta}$$

where $m$ is the revised mass.

### 3.3 The Electroshell

We use the results in the last section to find the generalised hierarchy, which applies to both the gravitational and electromagnetic interactions, whether terrestrial or celestial. We will call it *the electroshell* and follow the method used by Bell and Diaz [2004a]. We may translate equation (3.2.L) into our chosen form using equations (1.3.B) and (1.3.C),

$$Z_s^{(0)} e = \sqrt{2}\beta \tilde{m}_s^{(0)}, \quad Z_p^{(0)} e = 2\sqrt{2}\beta \tilde{m}_s^{(0)}, \quad (3.3.\text{A})$$
$$m^{(0)} = 2\sqrt{2}\beta\varepsilon \tilde{m}_s^{(0)}, \quad \tilde{m}_s^{(0)} = \tilde{M}_s, \quad v = v^{(0)} = v_S$$

where the bracketed superscript always indicates the level in the hierarchy. Our superscripts indicate that we shall change the velocity, $v^{(0)}$, the mass of the particle, $m^{(0)}$, and $\tilde{m}_s^{(0)}$ with the level but will keep $\varepsilon$ fixed, that $\beta$ is constant and that we change $Z_s^{(0)}$ and $Z_p^{(0)}$ rather than the charge, $e$. $r$ does not vary with the level either. We will set,

$$v^{(1)} = v_g \qquad (3.3.\text{B})$$

for conformity with our previous convention [Bell and Diaz 2004a]. The Dirac energy for level zero is,

$$m_s = m_s^{(1)} = m^{(0)}\sqrt{1-v_S^2} = 2\sqrt{2}\beta\varepsilon\, \tilde{m}_s^{(0)}\sqrt{1-v_S^2} \qquad (3.3.\text{C})$$

where the second equality comes from equations (2.2.A) and the last from equations (3.3.A). $m_s$ is the origin of the source charge for the interaction one level higher, and the source charge itself, $\tilde{m}_s^{(1)}$, is given by [Bell and Diaz 2004a],

$$m_s = m_s^{(1)} = \tilde{m}_s^{(1)}\sqrt{1-v_g^2} \qquad (3.3.\text{D})$$

Applying equation (3.3.D) to level zero as well and using equation (3.3.C),

$$m_s^{(1)} = 2\sqrt{2}\beta\varepsilon\, m_s^{(0)} \tag{3.3.E}$$

adding Bohr's first equation at levels zero and one using equations (3.2.L), (3.3.A) and (3.3.D),

$$v_g^2 = 2\sqrt{2}\beta\varepsilon\, v_s^2 \tag{3.3.F}$$

We find by inspection of the copyshell hierarchy in table 3 that,

$$2 \to 2\sqrt{2}\beta\varepsilon \tag{3.3.G}$$

reproduces the relation between the velocities of the particle at two adjacent levels. Our previous study [Bell and Diaz 2004a] and equation (3.2.A) shows us that we may assign the description using curvature to the first level of the hierarchy, while the description using motion is assigned to level zero, if we set,

$$2\sqrt{2}\beta\varepsilon = \frac{1}{\chi^2} \tag{3.3.H}$$

From equations (1.3.D), (3.1.A) and (3.1.C) and relation (3.3.G) we know that this factor must be $1/\tilde{\alpha}^2$, leading to,

$$\chi = \tilde{\alpha}, \ \ \varepsilon = \frac{1}{2\tilde{\alpha}^2}, \ \ \beta = \frac{1}{\sqrt{2}}, \ \ 2\sqrt{2}\beta\varepsilon = \frac{1}{\tilde{\alpha}^2} \tag{3.3.I}$$

from equation (3.2.C). Equations (3.3.A) become

$$Z_s^{(0)} e = \tilde{m}_s^{(0)}, \ \ Z_p^{(0)} e = 2\tilde{m}_s^{(0)},$$
$$m^{(0)} = \tilde{m}_s^{(0)}/\tilde{\alpha}^2, \ \ m_s^{(0)} = \tilde{M}_s, \ \ v = v^{(0)} = v_S \tag{3.3.J}$$

The appropriate value of the generalised Planck's constant, $h_g$, is calculated from the second of Bohr's equations, (1.3.C). It varies with the level we decide to quantise, which is the level at which the observer resides. Equations (3.2.L) quantise the hierarchy at level zero. This results in Bohr's equations for electromagnetism for that level, provided we set,

$$v_s = \tilde{\alpha} w = \frac{4\pi (\tilde{m}_s^{(0)})^2}{h_g} \qquad (3.3.K)$$

from equations (1.3.B), (1.3.C) and (3.3.J). It is at this point we could specialise to an hierarchy of the same form as the unishell, rather than the copyshell, by setting $w = \tilde{\alpha}$, but in that case table 4 immediately below could not be framed in the same way as some terms do not remain finite. This occurs when we include level two, where the velocity would equal the speed of light. We discuss other methods of avoiding the difficulty below.

**Table 4.**

**Part of the electroshell hierarchy**

| Level | -1 | 0 | 1 | 2 | 3 |
|---|---|---|---|---|---|
| $m_s^{(r)}$ $= E_e^{(r-1)}$ | $m_s\tilde{\alpha}^4$ $= E_e^{(-2)}$ | $m_s\tilde{\alpha}^2$ $= E_e^{(-1)}$ | $m_s$ $= E_e^{(0)}$ | $m_s/\tilde{\alpha}^2$ $= E_e^{(1)}$ | $m_s/\tilde{\alpha}^4$ $= E_e^{(2)}$ |
| **Source Charge** | $\dfrac{m_s\tilde{\alpha}^4}{\sqrt{1-w^2\tilde{\alpha}^4}}$ | $\dfrac{m_s\tilde{\alpha}^2}{\sqrt{1-w^2\tilde{\alpha}^2}}$ | $\dfrac{m_s}{\sqrt{1-w^2}}$ | $\dfrac{m_s}{\tilde{\alpha}^2\sqrt{1-\dfrac{w^2}{\tilde{\alpha}^2}}}$ | $\dfrac{m_s}{\tilde{\alpha}^4\sqrt{1-\dfrac{w^2}{\tilde{\alpha}^4}}}$ |
| **Particle Charge** | $\dfrac{2m_s\tilde{\alpha}^4}{\sqrt{1-w^2\tilde{\alpha}^4}}$ | $\dfrac{2m_s\tilde{\alpha}^2}{\sqrt{1-w^2\tilde{\alpha}^2}}$ | $\dfrac{2m_s}{\sqrt{1-w^2}}$ | $\dfrac{2m_s}{\tilde{\alpha}^2\sqrt{1-\dfrac{w^2}{\tilde{\alpha}^2}}}$ | $\dfrac{2m_s}{\tilde{\alpha}^4\sqrt{1-\dfrac{w^2}{\tilde{\alpha}^4}}}$ |
| **Particle Mass** $m_r$ | $\dfrac{m_s\tilde{\alpha}^2}{\sqrt{1-w^2\tilde{\alpha}^4}}$ | $\dfrac{m_s}{\sqrt{1-w^2\tilde{\alpha}^2}}$ | $\dfrac{m_s}{\tilde{\alpha}^2\sqrt{1-w^2}}$ | $\dfrac{m_s}{\tilde{\alpha}^4\sqrt{1-\dfrac{w^2}{\tilde{\alpha}^2}}}$ | $\dfrac{m_s}{\tilde{\alpha}^6\sqrt{1-\dfrac{w^2}{\tilde{\alpha}^4}}}$ |
| **Velocity** $v_r$ | $w\tilde{\alpha}^2$ | $w\tilde{\alpha} = v_S$ | $w = v_g$ | $\dfrac{w}{\tilde{\alpha}}$ | $\dfrac{w}{\tilde{\alpha}^2}$ |
| **Energy** $E_e^{(r)}/m_r$ | $\sqrt{1-w^2\tilde{\alpha}^4}$ | $\sqrt{1-w^2\tilde{\alpha}^2}$ | $\sqrt{1-w^2}$ | $\sqrt{1-\dfrac{w^2}{\tilde{\alpha}^2}}$ | $\sqrt{1-\dfrac{w^2}{\tilde{\alpha}^4}}$ |

The velocity, $v_r$, comes from equations (3.3.K) to find $v_S$, and (3.3.F) and (3.3.I) to iterate. $m_s$ is found in terms of $\tilde{m}_s^{(1)}$ in equation (3.3.D), $\tilde{m}_s^{(0)}$ is defined in equation (3.3.J) and equations (3.3.E) and (3.3.I) are used to iterate. Then the charge of both the source and particle come from equations (3.3.J). The mass of the particle, $m_r$, comes from equations (3.3.J). The Dirac energy, $E_e^{(r)}$, comes from equations (3.3.C). We have used the same notation in the tables as before [Bell and Diaz 2004a] wherever possible, so that the results may be directly compared.

### 3.4 Renormalisation Using the Electroshell and Copyshell

The renormalised energy for electromagnetism using table 4 at level zero is given by,

$$R_e^{(0)} = \frac{E_e^{(0)}}{(1+v_0^2)}(v_0^0 + v_0^2 + v_0^4 + v_0^6 \ldots) \tag{3.4.A}$$

$$= \frac{E_e^{(0)}}{m_s(1+v_0^2)}\sum_{r=1}^{-\infty}(-1)^{r-1}m_s^{(r)} = \frac{E_e^{(0)}}{m_s(1+v_0^2)^2}$$

with $w = i$, from table 4 and equations (2.1.A), (3.1.A) and (3.1.C). This is the same expression as we found for the unishell. Alternatively, we may suppose that $w$ is real but $\tilde{\alpha}$ imaginary,

$$\tilde{\alpha} \to i\tilde{\alpha} \tag{3.4.B}$$

use the scaling nature of the hierarchies and find for even $q$,

$$R_e^{(q)} = \frac{E_e^{(q)}}{m_s^{(q+1)}(1+v_q^2)}\sum_{r=q+1}^{-\infty}m_s^{(r)} = \frac{E_e^{(q)}}{m_s^{(q+1)}(1+v_q^2)^2} \tag{3.4.C}$$

The restriction picks the levels where the observer sees the temporal co-ordinates lie in the same direction, that is, we have particles rather than tachyons. Here we have found the renormalisation series for both electromagnetism and gravity. For the former we set $w = 1$, while for the latter we set $\tilde{\alpha} = i/\sqrt{2}$, with $w = 1/\sqrt{2}$ for the unishell or $w = v_g$ for the copyshell, as we may check from tables 1, 2, 3 and 4. We discovered earlier [Bell et al. 2004a], [Bell and Diaz 2004a], that an imaginary value for the velocity, $\tilde{\alpha}$, corresponded to charge conjugation. Converting from an attractive to a repulsive interaction also corresponds to a shift from the terrestrial to the celestial case or vice versa because the sign of the potential energy changes. This in its turn corresponds to the exchange of a Minkowski space-time to one with a Euclidean signature [Bell and Diaz 2004b] or vice versa. Charge conjugation also, in addition, corresponds to a shift from an electromagnetic interaction to a gravitational one or vice versa because like charges repel in the first and attract in the second, but here we do not usually see this as a change in signature of the space.

The tables all hold with the radius, $r$, and level of excitation, $n_\theta$, held constant. If we allow the radius to vary we may adjust the mass of the particle as we wish and still keep Bohr's equations, (1.3.B) and (1.3.C), valid and $h_g$ for any particular level constant. Changing $n_\theta$ will affect $\tilde{\alpha}$ and a table with the new value is required but $h_g$ does not alter if the level does not. This is because of our assumption that $j$ in the last of equations (3.2.L) changes conformably [Bell and Diaz 2004a]. $h_g = h$, Planck's constant, if we are quantising at level zero of the electroshell and describing electromagnetism.

## 4. DISCUSSION

Four interrelated threads have run through this paper. We have discussed in close conjunction, gravity, electromagnetism and their terrestrial and celestial variants. Really, we have done no more than show that many interactions have the same pattern. In this, the presence of the other in some shape or form has caused a mutual rotation. The interesting cases occur when more than one such rotation is involved. We suggest that this provides the impetus to form the hierarchical patterns we have just discussed. It is always necessary to choose a viewpoint, where the observer stands, and this is not simply a point in a space, curved or otherwise, but a level within a hierarchy. It is possible to move from viewpoint to another, and we discussed how this might be achieved. Ways of doing so in flat space are well known. For example, the observer might shift left or right, forward or backward or up or down in the familiar space of a room. It is equally possible to calculate how to do this in a hierarchy and when this is done some unfamiliar interaction may look like a familiar one. Electromagnetism, which is our best known interaction, can appear just like the unishell and with quantifiable differences like the gravitation of any body. Both may also have as adjuncts a field that travels infinitely fast, whatever the velocity of light.

**ACKNOWLEDGEMENTS**

One of us (Bell) would like to acknowledge the assistance of others including E. A. E. Bell. She would like to dedicate this paper to all our sons and daughters.